\documentclass[12pt]{iopart}

\renewcommand{\vec}[1]{{\bf #1}}

\usepackage[dvips]{epsfig}
\begin{document}
\jl{3}

\title[{\small\rm J. Phys. Condens. Matter {\bf 11}, 89 (1999)}]
      {Ferromagnetism and the temperature-dependent electronic structure 
      in thin Hubbard films}
\author{T. Herrmann and W. Nolting}
\address{Humboldt-Universit\"at zu Berlin, Institut f\"ur Physik,               
           Invalidenstr.\ 110, 10115 Berlin, Germany}
\date{\today}


\begin{abstract}
The magnetic behavior of thin ferromagnetic itinerant-electron 
films is investigated within the strongly correlated 
single-band Hubbard model. 
For its approximate solution we apply 
a generalization of the modified
alloy analogy (MAA) to deal with the modifications due to the
reduced translational symmetry. The theory is based on exact results in the
limit of strong Coulomb interaction which are important for a
reliable description of ferromagnetism. 
Within the MAA the actual type of the alloy analogy is determined
selfconsistently. The MAA allows, in particular, the investigation of
quasiparticle lifetime effects in the 
paramagnetic as well as the ferromagnetic phase. 
For thin fcc(100) and fcc(111) films the layer magnetizations are discussed as
a function of temperature as well as film thickness. 
The magnetization at the surface-layer is
found to be reduced compared to the inner layers. This reduction 
is stronger in fcc(100) than in fcc(111) films.
The magnetic behavior can be microscopically understood by means of the 
layer-dependent spectral density and the quasiparticle density of states.
The quasiparticle lifetime that corresponds to the width 
of the quasiparticle peaks in the spectral density is found to 
be strongly spin-  and temperature-dependent. 
\end{abstract}
\submitted
\maketitle
%
%
%
%
\section{Introduction}\label{sec_intro}
Remarkable advances in thin film technology have recently 
led to active interest in the nature of magnetism 
in ultrathin films, at surfaces
and multilayer structures. The influence of the reduced dimensionality on the
magnetic behavior of $3d$ transition metals has been extensively 
studied both experimentally 
\cite{All94,Bab96,Elm95,GBBS97,DTP+89,PKH90,FPA+97b} and theoretically
\cite{FF87,KMS94,AB97,LH96,SH97,RHGF95,GM94,PN97b,HPN98}. 
On the experimental side it was shown that ultrathin 
transition metal films can display
long range ferromagnetic order from a monolayer on \cite{DTP+89}. 
Although the Mermin-Wagner theorem \cite{MW66} 
requires the transition temperature to vanish
for perfectly isotropic two dimensional systems, it was shown 
theoretically that even a small amount of anisotropy may lead to magnetic 
order with a substantial transition temperature \cite{BM88,DVT96,SN98}. 
In real materials magnetic anisotropy is always present by virtue 
of either the dipole interaction or the spin-orbit coupling.

Theoretically, the $T=0\,$K properties of thin transition metal films have been
addressed by ab initio calculations 
within the density functional theory in the
local density approximation 
\cite{FF87,KMS94,AB97,LH96}.
However, these approaches are strictly based on a Stoner-type
model of  ferromagnetism and, therefore, treat electron
correlation effects which are responsible for the spontaneous magnetic order on
a low level. In addition they are restricted to ground state properties only.
To overcome this restriction, for example,   
a generalization of the fluctuating local moment method has been used
\cite{RHGF95}  to calculate the temperature-dependent electronic structure of
thin ferromagnetic films. However, the layer magnetizations at finite
temperatures and the magnetic short range order are needed as an input. In Ref.
\cite{SH97} magnetic phase transitions in thin films 
are investigated via a mapping of the ab initio 
results onto an effective Ising model.
Hasegawa calculates the finite temperature properties of thin 
Cu/Ni/Cu sandwiches
by use of the single-site spin-fluctuation theory \cite{Has87}.

For the understanding of the thermodynamical properties of thin film magnetism 
theoretical investigations on rather idealized model systems have proven to be
a good starting point. In this context several 
authors have focused on localized spin models
like the Heisenberg model \cite{HBCC72,EM91a,JDB92,SY92}.  
For example,  the mechanism that leads to the
experimentally observed temperature induced reorientation of the direction of
magnetization in thin Fe and  Ni films \cite{PKH90,FPA+97b} was 
investigated  in great detail \cite{HU97,JB98}. 
On the other hand it is by no means clear  to what extent the results 
obtained by localized spin models are applicable 
to transition metal films, where the
magnetically active electrons are itinerant.

The aim of the present paper is to study the interplay between strong electron 
correlations and the reduced translational symmetry due to the film geometry
within an itinerant electron model system. In particular we are interested in
the influence of the reduced dimensionality on spontaneous 
ferromagnetism and the spin-,
layer- and temperature-dependent electronic structure. 
For this purpose we restrict ourselves, at present, to the investigation of the
single-band Hubbard model \cite{Hub63}, which 
includes the minimum set of terms necessary  
for the description 
of itinerant-electron magnetism. 
The Hubbard model was originally introduced to explain band magnetism in
transition metals 
and has become a
standard model to study the essential physics of strongly correlated
electron systems over the years. 
It is clear that a realistic and quantitative description of ferromagnetism in
transition metals requires the inclusion of the degeneracy of the 3d-bands
\cite{Hub64a,VBH+97,HV98,MK98,NBDF89,HCO+94,DDP97}. 
Although the band-degeneracy is neglected in our model study, we believe
that a treatment of electron correlation effects well beyond
Hartree-Fock theory  will provide important insight into generic properties of
thin film ferromagnetism. For example, 
contrary to the expectation  on the basis 
of the well-known Stoner criterion,  
the magnetic order at the film surface may be reduced 
and less stable compared  to the inner layers 
if electron correlations are taken into 
account properly \cite{Has92}. 

Despite its apparent simplicity no general
solution exists until now for the Hubbard model.  
However, recently exact results have been obtained
by finite temperature quantum Monte Carlo calculations in the limit of infinite
dimensions \cite{Ulm98,VBH+97} which prove  the existence of 
ferromagnetic solutions  for
intermediate to strong Coulomb interaction $U$. 
In addition the decisive  importance of the
lattice geometry, i.e. the dispersion and distribution of spectral weight in
the non-interacting (Bloch) density of states (BDOS), 
on the magnetic stability was stressed by several 
authors \cite{VBH+97,Ulm98,Uhr96,HUM97,WBS+98,OPK97,HN97b}. 
A reasonable treatment of electron correlation effects  
led to an argument for the stability of ferromagnetism  
which is decisively more restrictive the well known Stoner criterion.
A BDOS with large spectral weight near one of the band edges is an essential
ingredient for ferromagnetism. The thermal stability of 
ferromagnetic solutions is favored by a 
strong asymmetry in the BDOS \cite{HUM97,WBS+98,PHWN98}.
This behavior of the BDOS is found, for example, in non-bipartite 
lattices like the fcc lattice.


Due to the broken translational symmetry even more
complications are introduced to the highly non-trivial many-body problem of the
Hubbard model. Thus we require an approximation scheme 
which is simple enough to allow for an extended study  of magnetic 
phase transitions and  electronic correlations  in  thin films. 
On the other hand it should be clearly beyond 
Hartree-Fock (Stoner) theory which has
been applied previously \cite{GM94}, since we believe a reasonable
treatment of electron correlation effects  
to be vital for a proper description 
of ferromagnetism especially for non-zero temperatures.

In this context interpolating theories which are essentially based on exact
results  
obtained by the $1/U$
perturbation theory first introduced by Harris and Lange \cite{HL67,EO94} 
have proven to be a good starting point \cite{PHWN98}. 
A theory that reproduces the 
rigorous strong coupling
results in a conceptual clear and straightforward manner 
is given by the spectral density approach (SDA) 
which has been discussed with respect to spontaneous magnetic order
for various three-dimensional \cite{NB89,HN97a,HN97b} as well as 
infinite dimensional \cite{HN97b,PHWN98} lattices.  
A similar approach applied to
a multiband Hubbard model led to surprisingly accurate results for the magnetic
key quantities of the prototype band ferromagnets Fe, Co, and Ni
\cite{NBDF89}.  A generalization of the SDA to systems with reduced
translational symmetry has recently  been given in  Refs. \cite{PN97b,HPN98},
which led, for example, to the description of the temperature-driven
reorientation transition within an itinerant-electron film \cite{HPN98}. 
However,  a severe limitation of the SDA results from the fact that 
quasiparticle damping is neglected completely. 
To tackle this problem a modified alloy analogy has been 
proposed \cite{HN96,NH98} 
which is closely related to the SDA but includes
quasiparticle damping effects 
in a natural way.
For bulk systems it was found that the
magnetic region in the phase diagram is significantly reduced by the inclusion
of damping effects. By comparison \cite{PHWN98,PHN98} 
with exact results for the fcc lattice in
the limit of infinite dimension and intermediate Coulomb interaction 
it is clear that  the Curie temperatures 
are somewhat overestimated within the MAA. 
However, the qualitative behavior of the ferromagnetic
solutions and in particular the dependence of the Curie temperature 
on the  band occupation is found to be in good agreement with the exact
results.  

In the present work we want to apply the MAA to systems with 
reduced translational symmetry.  
For this purpose the paper is
organized in the following way: 
In the next section we will give a short introduction to
the underlying many-body problem. 
The concept of an alloy analogy for the Hubbard
film  is developed in Sect.~\ref{sec_alloy}. In Sect.~\ref{sec_maa} 
we will generalize
the MAA to systems with reduced translational symmetry. 
The results of the numerical evaluations will be discussed
in  Sect.~\ref{sec_results} in terms of 
temperature- and layer-dependent magnetizations,
the quasiparticle bandstructure and 
the quasiparticle densities of states. We will
end with a conclusion in Sect.~\ref{sec_conclusion}.

\section{The many-body problem of the Hubbard film}\label{sec_many_body}
Let us first introduce the notation used to deal with the film geometry.
Each lattice vector of the film system is decomposed 
into two parts according to:
\begin{equation}
    \vec{R}_{i\alpha}=\vec{R}_i+\vec{r}_\alpha.
\end{equation}      
$\vec{R}_i$ denotes a lattice vector of the underlying two-dimensional
Bravais lattice with $N$ sites. 
To each of theses lattice sites there is associated a $d$-atom basis 
$\vec{r}_\alpha$ ($\alpha=1,\dots,d$) 
which refers to the $d$ layers of the film.
The same labeling with Latin and Greek indices applies  for all  
quantities related to the film geometry.
Within each layer we assume translational invariance. Then a 
two-dimensional Fourier transformation with respect to the 
Bravais lattice can be applied.

Using this notation the Hamiltonian for the single-band Hubbard film reads: 
\begin{equation} \label{hub_op}
        {\cal H}=\sum_{i,j,\alpha,\beta,\sigma}
	(T_{ij}^{\alpha\beta}-\mu\delta_{ij}^{\alpha\beta})
	c_{i\alpha\sigma}^{\dagger}c_{j\beta\sigma}+
	\frac{U}{2}\sum_{i,\alpha,\sigma}n_{i\alpha\sigma} n_{i\alpha-\sigma}.
\end{equation}
Here $c_{i\alpha\sigma}$ ($c_{i\alpha\sigma}^{\dagger}$) stands for 
the annihilation (creation) operator of an electron with spin $\sigma$ 
at the lattice site $\vec{R}_{i\alpha}$, 
$n_{i\alpha\sigma}=c_{i\alpha\sigma}^{\dagger}c_{i\alpha\sigma}$ 
is the number operator. 
$U$ denotes the on-site Coulomb matrix element and $\mu$ 
the chemical potential.
$T_{ij}^{\alpha\beta}$ is the hopping integral
between the lattice sites $\vec{R}_{i\alpha}$ and $\vec{R}_{j\beta}$. 
A two-dimensional Fourier transformation yields the corresponding dispersions
\begin{equation}
\label{hopping}
   T_\vec{k}^{\alpha\beta} =\frac{1}{N}\sum_{ij} T_{ij}^{\alpha\beta} 
                            e^{-i\vec{k}(\vec{R}_{i}-\vec{R}_{j})}.
\end{equation}
Here and in the following $\vec{k}$ denotes a wave-vector from the
underlying two-dimensional (surface) Brillouin zone.
Further we define 
$T_{0\alpha}=T_{ii}^{\alpha\alpha}=
\frac{1}{N}\sum_{\vec{k}}T_{\vec{k}}^{\alpha\alpha}=\textrm{const.}$ 
which gives the center of gravity of the $\alpha$-th layer in the BDOS. 

The basic quantity to be calculated is the retarded single-electron 
Green function 
\begin{equation}\label{green}
       G_{ij\sigma}^{\alpha\beta}(E)=\langle\langle c_{i\alpha\sigma};
       c_{j\beta\sigma}^{\dagger}\rangle\rangle_E.
\end{equation}
From $G_{ij\sigma}^{\alpha\beta}(E)$ we can obtain
all relevant information on the system. 
After a two-dimensional Fourier transformation 
one obtains from $G_{ij\sigma}^{\alpha\beta}(E)$ 
the spectral density
\begin{equation}
   S_{\vec{k}\sigma}^{\alpha\beta}(E)=-\frac{1}{\pi} 
   \textrm{Im} G_{\vec{k}\sigma}^{\alpha\beta}(E),
\end{equation}
which represents the bare lineshape of a (direct, inverse) photoemission
experiment.  
The diagonal elements of the Green function determine the 
spin- and layer-dependent quasiparticle density of states (QDOS):
\begin{equation}\label{eq:qdos}
        \rho_{\alpha\sigma}(E)=
        \frac{1}{N}\sum_{\vec{k}}S_{\vec{k}\sigma}^{\alpha\alpha}(E-\mu)
        =-\frac{1}{\pi} \textrm{Im}G_{ii\sigma}^{\alpha\alpha}(E-\mu).
\end{equation}
Via an energy integration one immediately gets from 
$\rho_{\alpha\sigma}(E)$ the band occupations
\begin{equation}\label{nas}
   n_{\alpha\sigma}\equiv\langle n_{i\alpha\sigma}\rangle=
   \int\limits_{-\infty}^{\infty} dE f_{-}(E) \rho_{\alpha\sigma}(E).
\end{equation}
$\langle\dots\rangle$ denotes the grand-canonical average and 
$f_{-}(E)$ is the Fermi function.
Here the site index $i$ has been omitted due to the assumed translational 
invariance within the layers.
Ferromagnetism is indicated by a spin-asymmetry in the band occupations
$n_{\alpha\sigma}$ leading to non-zero layer magnetizations 
$m_\alpha=n_{\alpha\uparrow}-n_{\alpha\downarrow}$.
The mean band occupation $n$ and the mean magnetization $m$ are given by
$n=\frac{1}{d}\sum_{\alpha\sigma}n_{\alpha\sigma}$ and
$m=\frac{1}{d}\sum_{\alpha}m_\alpha$, respectively.

The equation of motion for the single-electron Green function reads:
\begin{equation} \label{eq_motion}
\sum_{l\gamma}\left[(E+\mu)\delta_{il}^{\alpha\gamma}-T_{il}^{\alpha\gamma} 
          -\Sigma_{il\sigma}^{\alpha\gamma}(E)\right]
          G_{lj\sigma}^{\gamma\beta}(E) =\hbar\delta_{ij}^{\alpha\beta}.
\end{equation}
Here we  have introduced the electronic self-energy
$\Sigma_{ij\sigma}^{\alpha\beta}(E)$ which incorporates all effects of electron
correlations.

For later use we want to define the moments of the Green function 
\begin{equation}\label{mom1}
    M_{ij\sigma}^{(m)\alpha\beta}=-\frac{1}{\pi}\textrm{Im}
        \int\limits_{-\infty}^{\infty}\! dE \, E^{m} \,G_{ij\sigma}^{\alpha\beta}(E).
\end{equation}
The usefulness of the moments $M_{ij\sigma}^{(m)\alpha\beta}$ ($m=0,1,2,\dots$)
results from the fact that an alternative but equivalent representation can be
derived by use of the Heisenberg representation of the creation and annihilation
operators. Thus $M_{ij\sigma}^{(m)\alpha\beta}$ can be calculated up to the
desired order $m$ directly from the 
Hamiltonian (\ref{hub_op}) itself \cite{NB89,PN96}:
\begin{equation}\label{mom2}
    M_{ij\sigma}^{(m)\alpha\beta}=
    \bigg\langle\!\Big[
                       \underbrace{\big[...
	               [c_{i\alpha\sigma},{\cal H}]_{\!-}... ,  
                       {\cal H}\big]_{\!-}}_{m-\textrm{times}},
                        c_{j\beta\sigma}^{\dagger}
                  \Big]_{+}\!\bigg\rangle.
\end{equation}
Here $[\dots,\dots ]_{-(+)}$ denotes the commutator (anticommutator).
Eqs. (\ref{mom1}) and (\ref{mom2}) impose rigorous sum rules on the Green
function and the self-energy which have been
recognized to state important guidelines 
when constructing approximate solutions for
the Hubbard model \cite{PHWN98}. 
For example, the high energy expansion of the Green function 
is directly determined by the moments $M_{ij\sigma}^{(m)\alpha\beta}$.
It has been shown \cite{PHWN98} that 
the sum rules are  especially important in the limit of
strong Coulomb interaction:  
Being consistent with the sum rules up to the order $m=3$ 
states a necessary condition in order to
reproduce the exact results of the 
$1/U$-perturbation theory \cite{HL67,EO94}. 
Furthermore, the $m=3$ sum 
rule turns out to be of particular importance what
concerns the stability  of ferromagnetic 
solutions in the Hubbard model \cite{PHWN98}.

The sum rules up to order $m=3$ will be exploited in Sect.~\ref{sec_maa} 
for the construction of a modified alloy analogy (MAA) to the Hubbard film. 
First we want to introduce the concept of the alloy analogy approach 
for systems with reduced translational symmetry.

\section{The alloy analogy concept for the Hubbard film}\label{sec_alloy}
The main idea of the conventional alloy analogy  
approach \cite{Hub64b} is to consider, 
for the moment, the $-\sigma$-electrons to
be ``frozen'' and to be randomly distributed over the sites of the lattice.
Then a propagating $\sigma$-electron encounters a situation which is equivalent
to a fictitious alloy: At empty lattice sites it finds the atomic energy
$E_{1\sigma}$, at sites with a $-\sigma$-electron present the atomic energy
$E_{2\sigma}$. These energy levels are randomly distributed over the lattice
with concentrations $x_{1\sigma}$ and $x_{2\sigma}$ which correspond to the
probabilities for the $\sigma$-electron to meet these
respective situations.
Note that at this point it is not at all clear what choice of the energy levels and
concentrations gives the best approximation for the initial Hamiltonian.
However, an ``optimal'' choice of the alloy analogy parameters
should by some means 
account for the itineracy of the $-\sigma$-electrons (see Sect.~\ref{sec_maa}).
In the present film system the energy levels and concentrations may, in
addition, exhibit a layer-dependence. 
Thus the alloy analogy for the Hubbard film is
described by $4\cdot d$ a priori unknown parameters
\begin{equation}\label{ap}
    E_{1\sigma}^{(\alpha)},\,\,x_{1\sigma}^{(\alpha)},\,\, 
    E_{2\sigma}^{(\alpha)},\,\,x_{2\sigma}^{(\alpha)} 
     \qquad \alpha=1,\,\dots,\,d
\end{equation}

For the solution of the fictitious alloy problem given by (\ref{ap}) the 
coherent potential approximation (CPA) \cite{VKE68} 
provides a well known method.
The CPA has been realized to be the rigorous solution 
of the alloy problem in the
limit of infinite dimensions \cite{VV92} 
where the single-site aspect used in the derivation
of the CPA  becomes exact. In this sense the CPA can be termed to be the best
single-site approximation to the alloy problem. 
Due to the single-site aspect and the assumed translational invariance
within the layers we have 
$\Sigma_{ij\sigma}^{\alpha\beta}(E)=
\delta_{ij}^{\alpha\beta}\Sigma_{\alpha\sigma}(E)$. 
The implicit CPA equation \cite{VKE68} for the self-energy  
is readily formulated 
via an effective medium approach 
similar to the one discussed in Ref.~\cite{PN96}:
\begin{equation} \label{cpa}
   0=\sum_{p=1,2} x_{p\sigma}^{(\alpha)}
   \frac{E_{p\sigma}^{(\alpha)}-\Sigma_{\alpha\sigma}(E)-T_{0\alpha}}
   {1-\frac{1}{\hbar}G_{ii\sigma}^{\alpha\alpha}
     [E_{p\sigma}^{(\alpha)}-\Sigma_{\alpha\sigma}(E)-T_{0\alpha}]}.
\end{equation}
In addition the self-energy appears implicitly in the expression for the local
Green function which is given by matrix inversion from (\ref{eq_motion}) after
applying a two-dimensional Fourier transformation:
\begin{equation} \label{local_g}
  G_{ii\sigma}^{\alpha\beta}(E) =\frac{\hbar}{N}\!\sum_{\vec{k}}\!\!
  \mbox{\footnotesize
  $\displaystyle
   \left(\begin{array}{ccc}
             \!E\!+\!\mu-T_{\vec{k}}^{11}-\Sigma_{1\sigma}(E)&
                                          -T_{\vec{k}}^{12}&\dots\\
               -T_{\vec{k}}^{21}&\!\!\!\!\!E\!+\!\mu-T_{\vec{k}}^{22}
                                        -\Sigma_{2\sigma}(E)&\ddots\\
                         \vdots&\ddots&\ddots\\
  \end{array}\right)$}^{-1}_{\alpha\beta} \\[0.5cm]
\end{equation}
Eqs. (\ref{cpa}) and (\ref{local_g}) have to be solved selfconsistently to
obtain $G_{ii\sigma}^{\alpha\beta}(E)$ and $\Sigma_{\alpha\sigma}(E)$.

\section{The modified alloy analogy}\label{sec_maa}

Up to now nothing has been said about the actual choice of the atomic energies 
$E_{p\sigma}^{(\alpha)}$ and the corresponding concentrations
$x_{p\sigma}^{(\alpha)}$  ($p=1,2$).
In the conventional alloy analogy (AA) 
\cite{Hub64b}  the alloy parameters (\ref{ap}) 
are taken from the zero-bandwidth limit 
which directly corresponds to the assumption of strictly  ``frozen''
$-\sigma$-electrons:
\begin{eqnarray}
 &\tilde E_{1\sigma}^{(\alpha)}=T_{0\alpha}\qquad  
  &\tilde E_{2\sigma}^{(\alpha)}=T_{0\alpha}+U \nonumber\\
 &\tilde x_{1\sigma}^{(\alpha)}=1-n_{\alpha-\sigma}\qquad  
  &\tilde x_{2\sigma}^{(\alpha)}=n_{\alpha-\sigma}.\label{aa}
\end{eqnarray}
However, it was soon realized that the AA is not able to describe itinerant
ferromagnetism\cite{SD75}. This is closely 
related to the fact that the energy levels
$\tilde E_{p\sigma}^{(\alpha)}$ are rigid and, 
in particular, spin-independent quantities within the AA. Further it is known
\cite{HN96,NH98,PHN98} that the AA fulfills 
the sum rules (\ref{mom1}), (\ref{mom2})  up to the
order $m=2$ only and fails to reproduce the 
correct strong coupling behavior. Note
that within the AA the energy levels $\tilde E_{p\sigma}^{(\alpha)}$ 
are layer-independent (for uniform $T_{0\alpha}$) 
which  is  a crude approximation since  a possible layer-dependence in the
quasiparticle spectrum is suppressed almost completely.   

The basic idea  of the MAA is 
to exploit the information provided by the non-trivial but exact results in the
limit of strong Coulomb interaction ($U/t\gg1$) \cite{HL67,EO94} to determine
the energy levels and concentrations (\ref{ap}).
This can most elegantly be achieved  by imposing the 
sum rules (\ref{mom1}),
(\ref{mom2}) on the CPA equation (\ref{cpa}) \cite{PHN98,PHWN98}: 
By inserting the high energy expansion of the self-energy 
$\Sigma_{\alpha\sigma}(E)$ and the local Green function
$G_{ii}^{\alpha\alpha}(E)$, which are determined by the sum rules, 
the CPA equation (\ref{cpa}) can be expanded in powers of $1/E$. 
Taking into
account the sum rules up to the order  $m=3$ unambiguously determines the
parameters $E_{p\sigma}^{(\alpha)}$, $x_{p\sigma}^{(\alpha)}$.
Then the exact strong coupling results 
are reproduced automatically \cite{PHWN98}. Note that due to the single site
aspect of the CPA only the local terms of the $1/U$ perturbation theory are
reproduced. 
On the other hand  the MAA is not restricted solely to the  strong
coupling  limit but is also applicable for intermediate 
interaction strengths where it has
an interpolating character \cite{HN96,NH98}.
Following this procedure yields the energy levels and concentrations of the MAA
for the Hubbard film:
\begin{eqnarray}
    E_{p\sigma}^{(\alpha)}&=& 
    {1\over 2} \bigg[T_{0\alpha}+U+B_{\alpha-\sigma}(\pm)^p\nonumber\\*
    &&\sqrt{(U+B_{\alpha-\sigma}-T_{0\alpha})^2+
    4 U n_{\alpha-\sigma}(T_{0\alpha}-B_{\alpha-\sigma})}\bigg],
    \nonumber\\*
    x_{1\sigma}^{(\alpha)}&=&
    \frac{B_{\alpha-\sigma}+U(1- n_{\alpha-\sigma})-E_{1\sigma}^{(\alpha)}}
    {E_{2\sigma}^{(\alpha)}-E_{1\sigma}^{(\alpha)}}\label{maa}\\
    x_{2\sigma}^{(\alpha)}&=&1-x_{1\sigma}^{(\alpha)}\nonumber
\end{eqnarray}
An alternative derivation of the MAA  for bulk systems which is based 
on physical arguments can be found in Refs. \cite{HN96,NH98}.
Note that the expressions for 
$E_{p\sigma}^{(\alpha)}$ and $x_{p\sigma}^{(\alpha)}$ in (\ref{maa}) 
are directly related to the position and the weight of 
the two poles of the spectral density
within the SDA. 
Eqs. (\ref{maa}) are obtained from the SDA results \cite{HN96,NH98} if the 
electron dispersion is replaced by the
center of gravity of the non-interacting band.  
The energy levels and concentrations (\ref{maa}) 
are not only dependent on the model
parameters $T_{0\alpha}$ and $U$ but also on the band occupations
$n_{\alpha-\sigma}$ and the so-called bandshift $B_{\alpha-\sigma}$. 
The bandshift
that is introduced via the fourth moment $M_{ij-\sigma}^{(3)\alpha\beta}$ 
consists of
higher correlation functions:
\begin{eqnarray}
   B_{\alpha-\sigma}&=&\,T_{0\alpha}\!+\!
   \frac{1}{n_{\alpha-\sigma}(1\!-\!n_{\alpha-\sigma})}
	\sum_{j,\beta}^{j\beta\neq i\alpha} 
	T_{ij}^{\alpha\beta}\langle c_{i\alpha-\sigma}^{\dagger} 
	c_{j\beta-\sigma}(2n_{i\alpha\sigma}\!-\!1)\rangle.
\end{eqnarray} 
Nevertheless  $B_{\alpha-\sigma}$ 
can exactly be calculated \cite{NB89,HN97a} by use of the local Green
function and the self-energy:
\begin{eqnarray}
   B_{\alpha-\sigma}&=&T_{0\alpha}+
   \frac{1}{n_{\alpha-\sigma}(1-n_{\alpha-\sigma})}
   \frac{1}{\hbar}\,\textrm{Im}\!\!\!\int\limits_{-\infty}^{+\infty}\!\!\!dE\,
   f_{-}(E)
   \left(\frac{2}{U}\Sigma_{\alpha-\sigma}(E-\mu)-1\right)
   \times\nonumber\\
   &&[(E-\Sigma_{\alpha-\sigma}(E-\mu)-T_{0\alpha})
   G_{ii-\sigma}^{\alpha\alpha}(E-\mu)-\hbar]
\label{bas}
\end{eqnarray}

In the strict zero-bandwidth limit $B_{\alpha-\sigma}$ is 
identical to $T_{0\alpha}$ and the  MAA (\ref{maa}) reduces 
to the conventional alloy analogy (\ref{aa}). 
However, as soon as the hopping is
switched on, the bandshift $B_{\alpha-\sigma}$, 
which is for strong Coulomb interaction
proportional to the kinetic energy of the $-\sigma$-electrons 
in the $\alpha$-th layer \cite{HN97a}, 
has to be calculated selfconsistently by iteration.
Thus, via (\ref{maa}) the type of the underlying alloy changes 
in each step of the iteration process. In this sense $B_{\alpha-\sigma}$
accounts for the itineracy of the $-\sigma$-electrons. 
In the paramagnetic phase there are only minor differences in the quasiparticle
spectrum between MAA and AA. 
However, the bandshift may get a real spin-dependence
for special parameter constellations.
Thus $B_{\alpha-\sigma}$ 
may generate and stabilize ferromagnetic solutions which are excluded within
the AA. It is worth to stress that the energy levels 
and concentrations are implicitly
temperature-dependent via $n_{\alpha-\sigma}$ and $B_{\alpha-\sigma}$ leading,
therefore, to a temperature dependent electronic structure.

The evaluation of the MAA requires the solution of two nested 
selfconsistency cycles. One starts
with an initial guess for the band occupations $n_{\alpha-\sigma}$ and the
bandshift $B_{\alpha-\sigma}$ which determine the 
energy levels and concentrations (\ref{maa}). Via
the CPA-equation (\ref{cpa}) and (\ref{local_g}) 
the corresponding self-energy and
Green function can be calculated selfconsistently. 
With this solution new values for $n_{\alpha-\sigma}$ and $B_{\alpha-\sigma}$
are obtained via (\ref{nas}) and (\ref{bas}). 
This procedure is iterated until convergence is achieved.  
For efficiency reasons the numerical evaluations 
of the integrals in (\ref{nas})  and (\ref{bas}) are performed via discrete
Matsubara sums on the imaginary energy axis \cite{PHN98}.
Only the spectral density and the quasiparticle density of states
are calculated on the real axis at the end of each 
selfconsistency procedure.

\section{Results and Discussion}\label{sec_results}

For the numerical evaluations we consider in the present work 
thin fcc films with an (100) as well as an (111) surface and a 
film thickness up to $d=15$.  
The hopping integral  between nearest neighbor sites is chosen to be 
uniform throughout the film and is set to $t=-0.25\,$eV. All other hopping
integrals as well as  $T_{0\alpha}$ are set to zero. 
For an fcc bulk system
this yields a total bandwidth $W^{\textrm{\scriptsize bulk}}=4\,$eV 
of the non-interacting system. 
Further, we keep the on-site Coulomb interaction fixed at
$U=50\,$eV which clearly refers to the strong coupling regime.
In all calculations the 
total band occupation is kept fixed at the representative value $n=1.6$. Bulk
calculations within the MAA have shown \cite{NH98} 
that for the fcc lattice ferromagnetic order 
is possible for all band occupations above half filling $n>1$.

\begin{table}[h]
\caption{Number of nearest neighbors (n.n.) for the fcc lattice as well as
the fcc(100) and the fcc(111) film structure. In addition the moments
$\Delta_\alpha^{(m)}$ (see Eq.~(19)) of the BDOS are given.
$\Delta_s^{(m)}$ refers to the surface layer, while
the moments of all other layers ($2\le\alpha\le d-1$) are  
equal to the respective bulk values $\Delta_b^{(m)}$.} 
\label{tab_one}
\begin{center}
{
\begin{tabular}{|lr|c|c|}
\hline
\multicolumn{2}{|c|}{bulk} & \multicolumn{2}{|c|}{fcc}\\ \hline
\multicolumn{2}{|c|}{n.n} & \multicolumn{2}{|c|}{12}\\
\multicolumn{2}{|c|}{$\Delta^{(2)}_b$} & \multicolumn{2}{|c|}{12}\\
\multicolumn{2}{|c|}{$\Delta^{(3)}_b$} & \multicolumn{2}{|c|}{-48}\\ 
\hline\hline
\multicolumn{2}{|c|}{film} & (100) & (111)\\[0.3ex] \hline
    & +1 & 4 & 3\\[0.3ex]
n.n.&  0 & 4 & 6\\[0.3ex]
    & -1 & 4 & 3\\[0.3ex] \hline
\multicolumn{2}{|c|}{$\Delta^{(2)}_s$} &   8 &   9\\
\multicolumn{2}{|c|}{$\Delta^{(3)}_s$} & -24 & -30\\  
\hline\hline
\multicolumn{2}{|c|}{$\Delta^{(2)}_s/\Delta^{(2)}_b$} &  0.667 &   0.750\\
\multicolumn{2}{|c|}{$\Delta^{(3)}_s/\Delta^{(3)}_b$} &  0.5   &   0.625\\ 
\hline    
\end{tabular}
}
\end{center}

\end{table}

\begin{figure}
	\centerline{\epsfig{figure=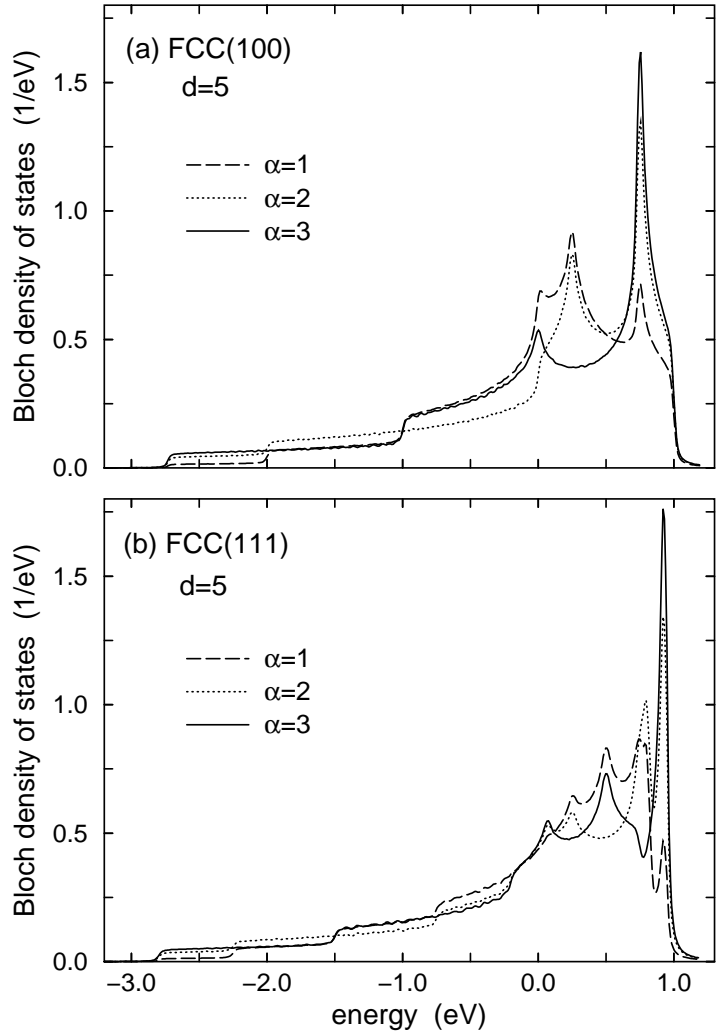,width=0.6\linewidth}}
	\caption{Layer-dependent Bloch density of states $\rho_{0\alpha}(E)$ of
	a five layer film for  (a) the fcc(100) and (b) the fcc(111) structure.
	$\alpha=1$ denotes the surface layer, $\alpha=3$ the central layer.
	Further parameters: $t=-0.25\,$eV, $T_{0\alpha}=0$.}
	\label{fig_bdos}
\end{figure}

For both film structures considered here, fcc(100) and fcc(111), all nearest
neighbors are placed in the same or in the  adjacent layer. 
The number of 
nearest neighbors in these two film geometries are given in Tab.~\ref{tab_one}.
The corresponding dispersions $T_{\vec{k}}^{\alpha\beta}$ 
can then be written as:
\begin{equation}
 T_{\vec{k}}^{\alpha\beta}=\left\{
              \begin{array}{ll}
              \,T_{0\alpha}+t\gamma_{||}(\vec{k}),\,\, & \alpha=\beta\\
              \,t\gamma_{\perp}(\vec{k}),\,\,& \alpha=\beta-1\\
              \,t(\gamma_{\perp}(\vec{k}))^{\star},\,\,&
              	   			       \alpha=\beta+1.\\
              \end{array}
              \right. 
\end{equation}
According to (\ref{hopping}) one gets 
\begin{eqnarray*}
\gamma_{||}^{(100)}(\vec{k})\,\,\,&=&
            2\Big[\cos(\frac{k_x+k_y}{2})+\cos(\frac{k_x-k_y}{2})\Big]\\
\gamma_{\perp}^{(100)}(\vec{k})&=&
            1+e^{-i\frac{k_x+k_y}{2}}+e^{-i\frac{k_x-k_y}{2}}
            +e^{-i k_x}
\end{eqnarray*}
for the fcc(100) film geometry and
\begin{eqnarray*}
\gamma_{||}^{(111)}(\vec{k})\,\,\,&=&
            2\Big[\cos(\sqrt{3/8}k_x+\sqrt{1/8}k_y)+
            \cos(\sqrt{3/8}k_x-\sqrt{1/8}k_y)+\nonumber\\
            &&\cos(\sqrt{1/2}k_y)\Big]\\
\gamma_{\perp}^{(111)}(\vec{k})&=&
            1+e^{-i(\sqrt{3/8}k_x+\sqrt{1/8}k_y)}
             +e^{-i(\sqrt{3/8}k_x-\sqrt{1/8}k_y)}       
\end{eqnarray*}
for fcc(111). Here the lattice constant is set to $a=1$. 
The layer-dependent Bloch density of states 
$\rho_{0\alpha}(E)=\rho_{\alpha\sigma}(E)|_{U=0}$ for a five layer film  is
plotted in Fig.~\ref{fig_bdos} for both film structures considered.
The BDOS is strongly asymmetric and shows a distinct layer
dependence. Considering the moments 
\begin{equation}\label{bdos_mom}
     \Delta_\alpha^{(m)}=\frac{1}{t^m}
     \int\limits_{-\infty}^{\infty} dE (E-T_{0\alpha})^m \rho_{0\alpha}(E) 
\end{equation}
of the BDOS yields  
that the variance $\Delta_\alpha^{(2)}$ as well as the
skewness $\Delta_\alpha^{(3)}$ are 
reduced at the surface layer compared to the
inner layers due to the reduced coordination 
number at the surface (see Tab.~\ref{tab_one}).

The charge distributions $n_\alpha$ as well as the 
layer magnetizations $m_\alpha$ are
determined by the selfconsistently calculated QDOS (\ref{eq:qdos}) via
(\ref{nas}).  The chemical potential $\mu$ and the band centers $T_{0\alpha}$
are assumed to be uniform throughout the film, allowing, therefore, for charge
transfer between the layers. However, in the actual calculation the difference
in the occupation numbers $n_\alpha$ turns out to be very small ($<3\%$).

\begin{figure}
	\centerline{\epsfig{figure=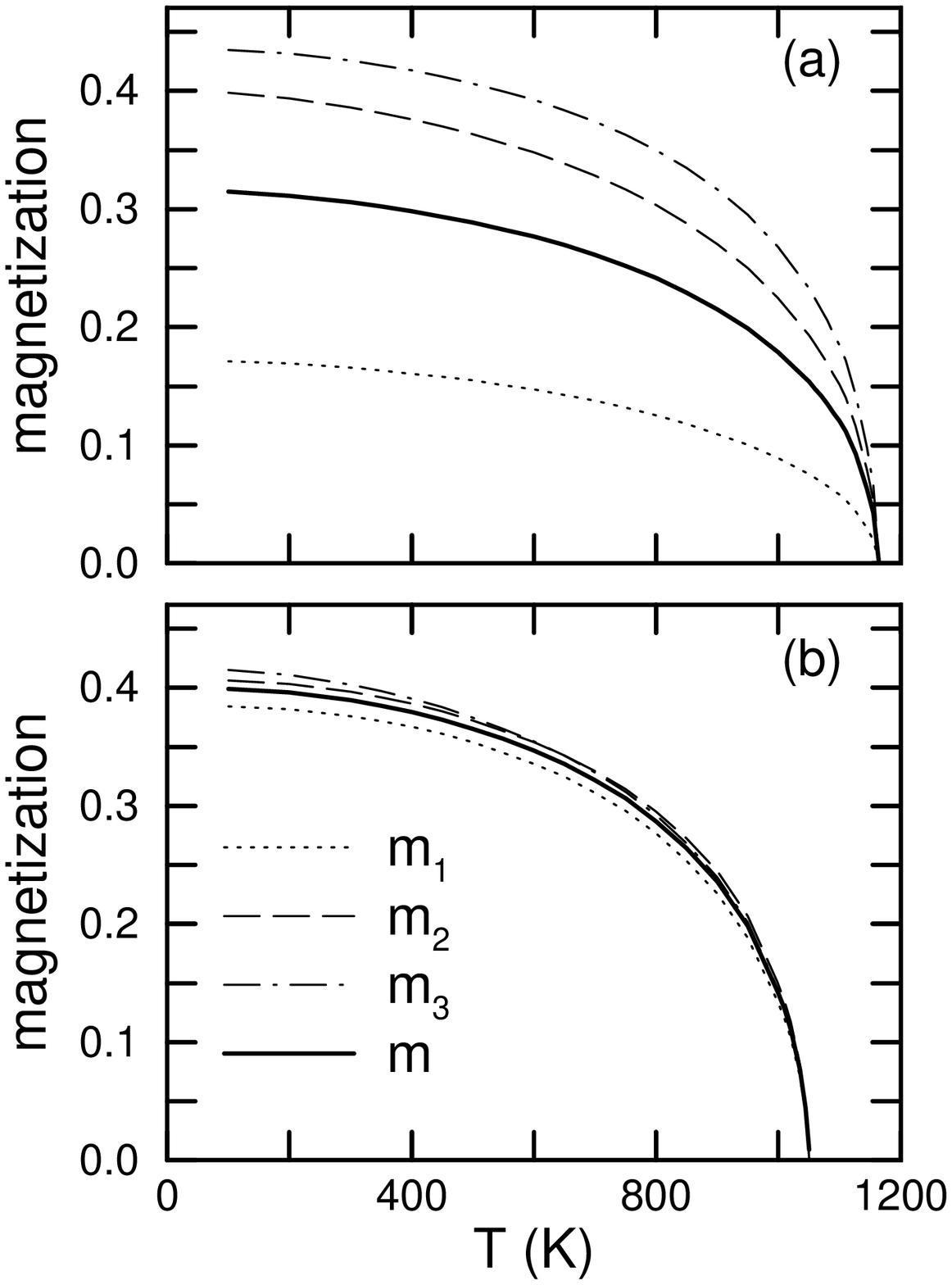,width=0.7\linewidth}}
	\caption{Layer magnetizations $m_\alpha$ as a function of temperature
	for (a) an fcc(100) and (b) an fcc(111) five layer film.
	$\alpha=1$ corresponds to the surface layer, $\alpha=3$ 
	to the central layer.
	Further parameters: $U=50\,$eV, $t=-0.25\,$eV, $n=1.6$.}
	\label{fig_m_t}
\end{figure}

The layer-dependent magnetizations $m_\alpha$ together with the mean
magnetization $m$ for a five layer film are plotted in Fig.~\ref{fig_m_t} 
as a function of temperature for both film structures. 
With respect to the overall shape the magnetization curves show the usual
Brillouin-type behavior.  
However, the surface magnetization is found to be 
reduced compared to the inner
layers for all temperatures. The reduction is 
particularly strong for the 
fcc(100) film geometry and 
leads to a non-saturated groundstate   
whereas the fcc(111) film is fully polarized at $T=0$.
The enhanced surface effects in the fcc(100) 
structure are related to the higher
percent of missing nearest neighbors at the surface 
layer which is $1/3$ for
fcc(100) and $1/4$ for fcc(111).

\begin{figure}
	\centerline{\epsfig{figure=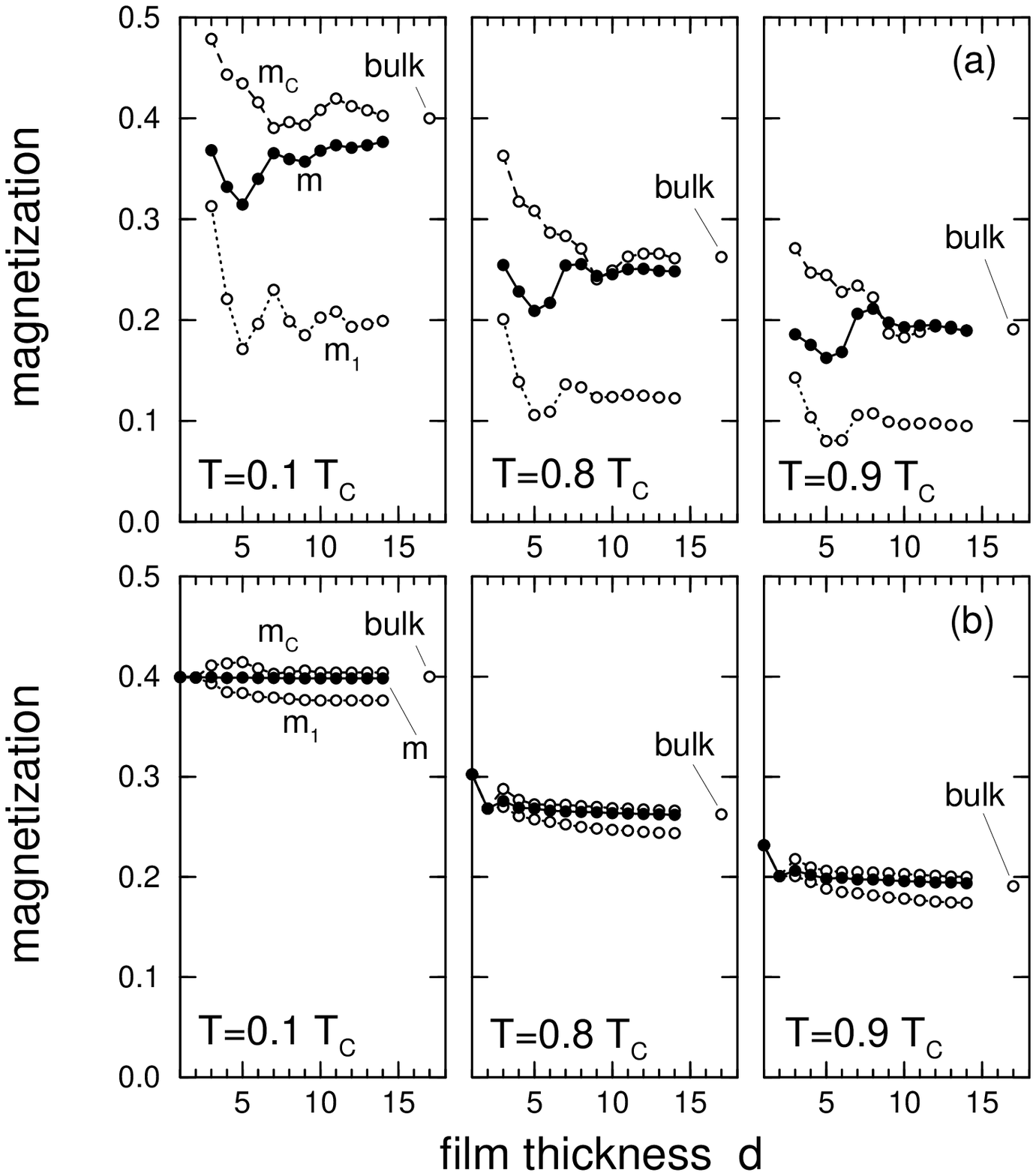,width=0.8\linewidth}}
	\caption{Surface-, center-, and mean magnetization ($m_1$, $m_c$, $m$)
	as a function of the film thickness $d$ for three different 
	temperatures $T=0.1\,T_C$,
	$T=0.8\,T_C$, and $T=0.9\,T_C$; (a) fcc(100); (b) fcc(111).
	Further parameters: $U=50\,$eV, $t=-0.25\,$eV, $n=1.6$.}
	\label{fig_m_d}
\end{figure}

We want to emphasize that the finding of a 
reduced surface magnetization cannot
be explained by the well-known Stoner criterion of ferromagnetism. 
Since the variance   
of the BDOS is reduced at the surface layer
($\Delta_s^{(2)}<\Delta_b^{(2)}$, see Tab.~\ref{tab_one})
due to the reduced coordination number
one might intuitively expect the magnetization at the
surface to be more robust than in the bulk. However, as discussed in
Sect.~\ref{sec_intro}, 
intensive investigations of strongly correlated electron systems 
well beyond Hartree-Fock (Stoner) theory clearly point out  the
importance of a large skewness $\Delta^{(3)}$ for the stability of
ferromagnetism \cite{HUM97,WBS+98,PHWN98}. Since the skewness of the BDOS 
is strongly reduced at the surface (see Tab.~\ref{tab_one}) 
this explains the trend of a reduced surface magnetization.
The above  argument can be checked by considering the BDOS of the surface layer 
(see. Fig.~\ref{fig_bdos}) as an input for an additional MAA calculation. 
Doing so we find that a ``separated'' surface layer 
would be ferromagnetic for the
fcc(111) but paramagnetic for the fcc(100) film structure.
In this sense the surface layer of an fcc(100) film is magnetized 
only because of the effective field induced by the ferromagnetically 
ordered inner layers.

The Curie temperature is found to be unique for the
whole film. Note, that although the  mean magnetization is reduced for the
fcc(100) film with respect to fcc(111) the corresponding Curie temperature is
enhanced ($T_C^{(100)}(d=5)=1140\,$K, $T_C^{(111)}(d=5)=1050\,$K). 
The inner layers that are fully polarized at
$T=0\,$K for both film structures appear to be magnetically 
more stable for fcc(100) compared to fcc(111). Again, this  trend can also be  
seen in an additional MAA calculation for the BDOS 
of the respective central layers. 
The Curie temperatures converge to the corresponding bulk value
($T_c^{\textrm{\scriptsize bulk}}=1050\,$K) for $d^{(111)}\approx 3$ 
and $d^{(100)}\approx 6$.

In Fig.~\ref{fig_m_d}  the surface-, center-, and mean magnetization ($m_1$,
$m_c$ and $m$) are shown
as a function of the film thickness. The surface magnetization is 
reduced compared to the mean magnetization. The
reduction is weak for fcc(111) films but very pronounced in the case of the
fcc(100) structure. This holds not only 
for thin films where some oscillations are
present due to the finite film thickness, but also extends to the limit
$d\rightarrow\infty$ where the two surfaces are well separated and do not
interact.  The oscillations as a function of $d$ which are present for the
fcc(100) structure get damped for higher temperatures. 
One can see from Fig.~\ref{fig_m_d} that the center 
layer magnetization $m_c$ for thick films ($d>10$) is in good agreement 
with the corresponding fcc bulk calculation
\cite{NH98}.

\begin{figure}
	\centerline{\epsfig{figure=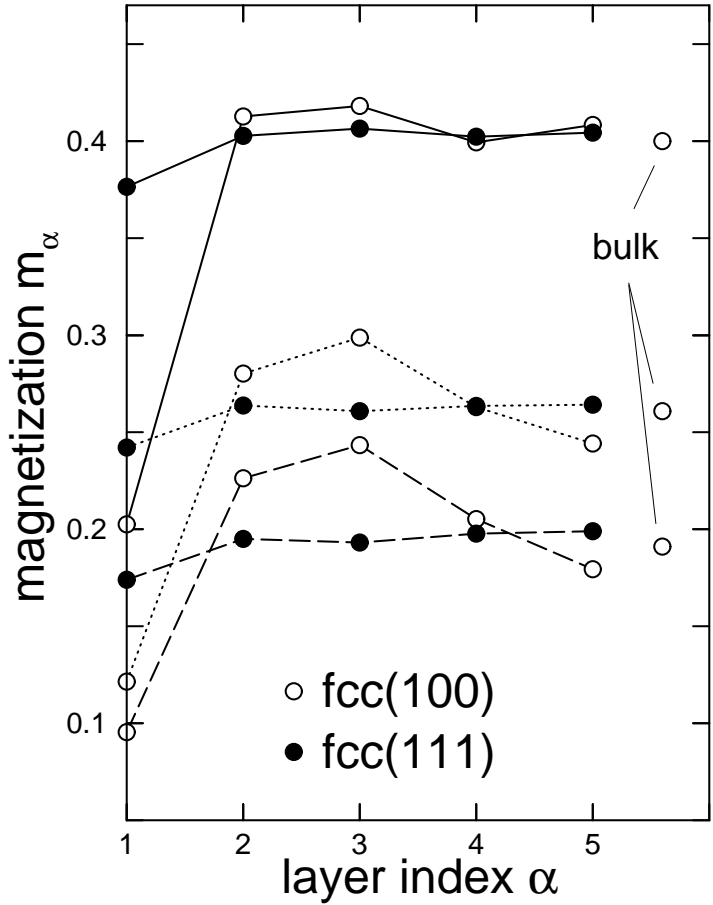,width=0.6\linewidth,angle=0}}
	\caption{Magnetization profile of a ten layer film 
	for three different temperatures $T=0.1\,T_C$ (solid line),
	$T=0.8\,T_C$ (dotted line), and $T=0.9\,T_C$ (dashed line).
	Further parameters: $U=50\,$eV, $t=-0.25\,$eV, $n=1.6$.}
	\label{fig_m_alpha}
\end{figure}

The magnetization profile for  both film geometries
is plotted for $d=10$ in Fig.~\ref{fig_m_alpha}. 
Here again the magnetizations of the fcc(100) film   
show a pronounced layer dependence while they are very close to the bulk value
from the second layer on in the case of the fcc(111) film geometry.
The magnetization profiles are similar to the ones obtained in 
\cite{Has87} for Cu/Ni/Cu sandwiches calculated within a single-site
spin-fluctuation theory. However, within the present approach the
deviation from the bulk magnetization  is
enhanced close to the Curie temperature for  the fcc(100)
structure (Fig.~\ref{fig_m_alpha}).
Note that a similar trend to a reduced surface magnetization is also found  
within localized spin models. However, 
for the uniform Heisenberg model without a layer-dependent anisotropy
contribution, the layer magnetizations necessarily 
increase {\em monotonously\/} from
the surface to the central layer \cite{HBCC72,SN98}.

\begin{figure}
 \centerline{\epsfig{figure=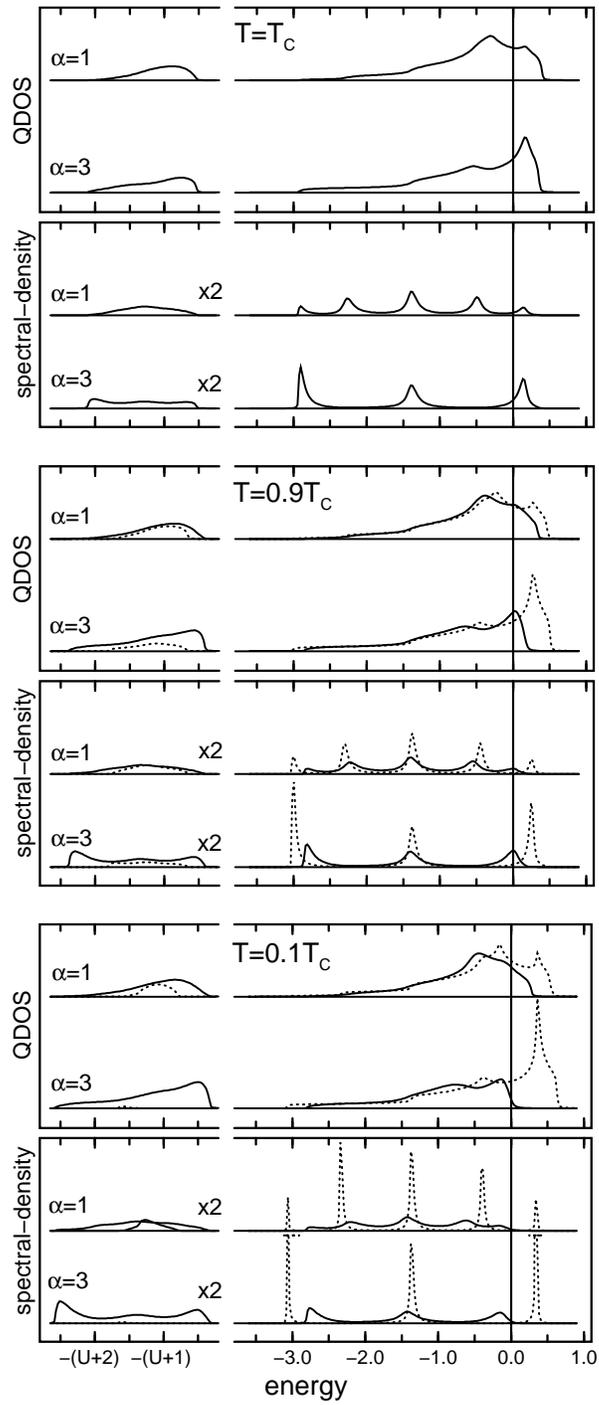,width=0.5\linewidth,angle=0}}
 \caption{Spectral-density at the gamma point ($\bar\Gamma$) and  
          quasiparticle density of states   
          of a five layer FCC(100) Hubbard film. 
          Only the surface layer ($\alpha=1$) and the central layer
          ($\alpha=3$) are shown.
          Solid lines: $\sigma=\uparrow$; 
          dotted lines: $\sigma=\downarrow$.
          The chemical potential is located at zero energy.
          Further parameters: $U=50\,$eV, $t=-0.25\,$eV, $n=1.6$.}
 \label{fig_sk_g_qdos}
\end{figure}

To understand the magnetic behavior on a  microscopic basis we 
will, in the following, discuss the temperature-dependent 
electronic structure of the thin film systems.
For a five layer fcc(100) film the 
spin- and layer-dependent spectral density at
the gamma $\bar\Gamma$ point and the quasiparticle density of states  are
plotted in Fig.~\ref{fig_sk_g_qdos}. There appear two correlation induced
band-splittings in the quasiparticle spectrum: 
Due to the strong Coulomb interaction the spectrum splits into a low and a high
energy subband (``Hubbard bands'') 
which are separated by an energy of the order  $U$. 
Besides this so-called ``Hubbard splitting'' that is present for 
all temperatures
there is an additional exchange splitting in majority
($\sigma=\uparrow$) and minority ($\sigma=\downarrow$) spin direction 
for temperatures below $T_C$.  
In the lower subband the
electron mainly hops over empty sites, while in the upper subband it hops
over lattice sites that are already occupied by another electron with opposite
spin. The corresponding weights of the 
subbands scale with the probability of the
realization of these two situations. 
In the strong coupling limit the scaling is given by
$(1-n_{\alpha-\sigma})$ and $n_{\alpha-\sigma}$ 
for the lower and upper subband, respectively. 
Since the total band occupation 
($n=1.6$) considered here is above half filling, 
the chemical potential $\mu$ is located in the
upper subband while the lower subband is completely filled.

\begin{figure}
  \centerline{\epsfig{figure=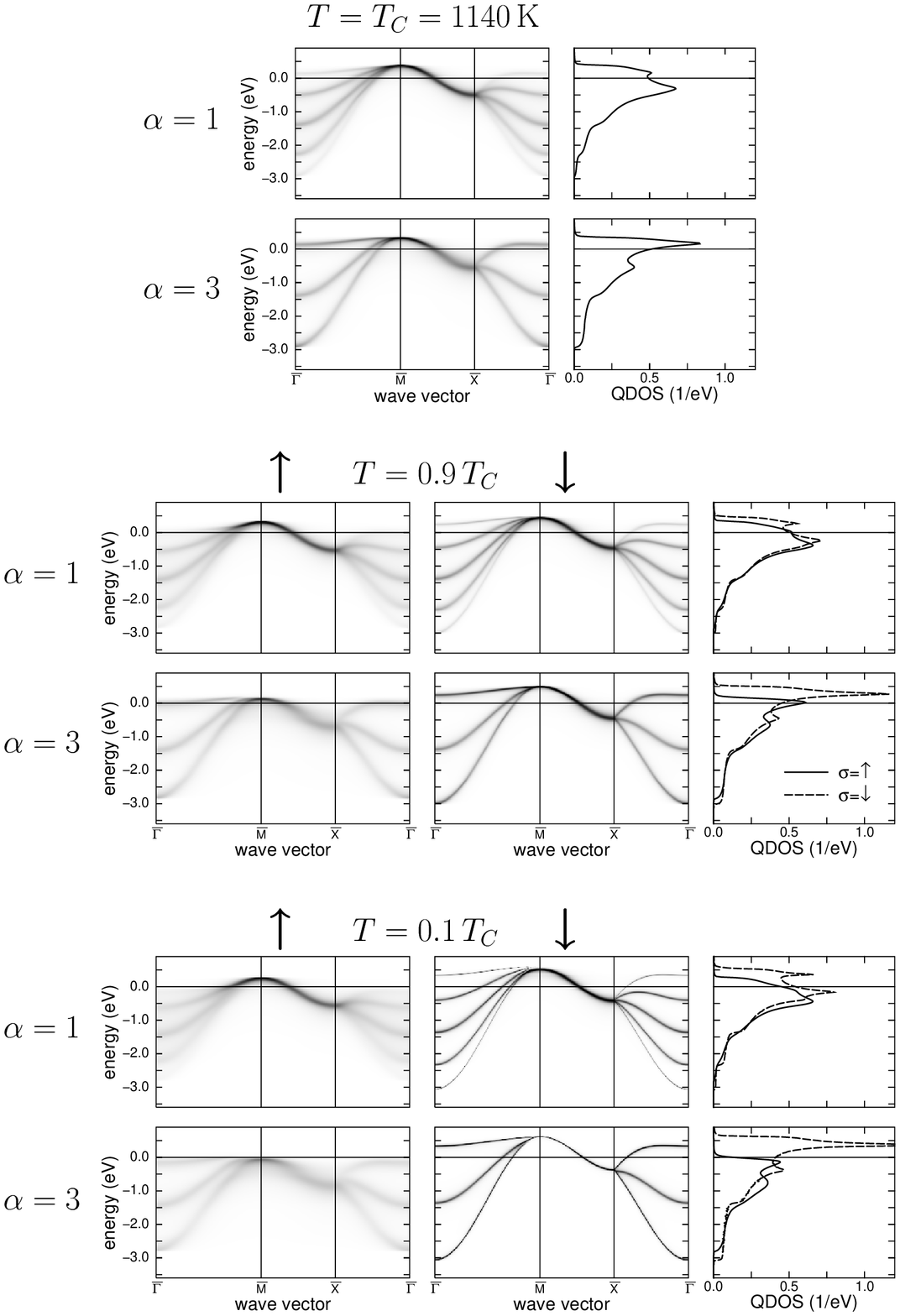,width=0.9\linewidth,angle=0}}
  \caption{Spin- and layer- dependent spectral density and 
  quasiparticle density of states of a five layer fcc(100) film
  for three different temperatures $T=0.1\,T_C$, $T=0.9\,T_C$, 
  and $T=T_C$. Only the upper Hubbard band is shown.
  $\alpha=1$: surface layer; $\alpha=3$: central layer.
  The chemical potential is located at zero energy.
  Further parameters: $U=50\,$eV, $t=-0.25\,$eV, $n=1.6$.}
  \label{fig_sk_qdos_100}
\end{figure}

\begin{figure}
  \centerline{\epsfig{figure=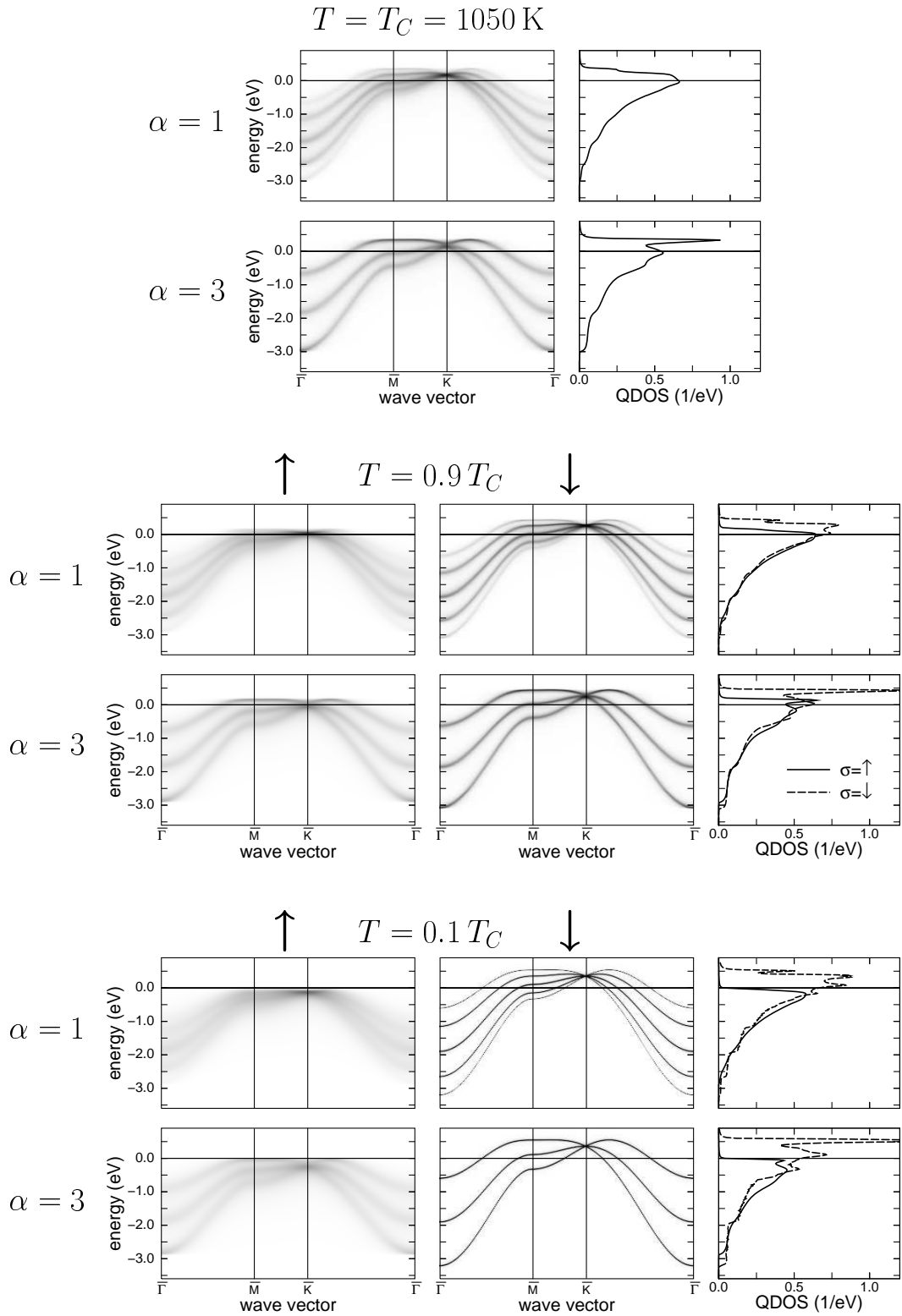,width=0.9\linewidth,angle=0}}
  \caption{Spin and layer dependent spectral density and quasiparticle
  density of states of a five layer fcc(111) film
  for three different temperatures $T=0.1\,T_C$, $T=0.9\,T_C$, 
  and $T=T_C$. Only the upper Hubbard band is shown.
  $\alpha=1$: surface layer; $\alpha=3$: central layer.
  The chemical potential is located at zero energy.
  Further parameters: $U=50\,$eV, $t=-0.25\,$eV, $n=1.6$.}
  \label{fig_sk_qdos_111}
\end{figure}

Starting from the Curie temperature the evolution 
of the quasiparticle spectrum with
decreasing temperature is dominated by two distinct correlation effects. 
Both are driven by an increasing
spin-asymmetry in the bandshift $B_{\alpha-\sigma}$.  Firstly 
the centers of gravity of the majority and minority subbands move apart 
with decreasing temperature (Stoner-type behavior). 
Secondly there is a strong spin-dependent transfer of spectral weight 
between the lower and the upper
subbands according to the above mentioned scaling, which results in spin- 
and temperature-dependent widths of the respective 
subbands.
This behavior can also be seen in detail in
Fig.~\ref{fig_sk_qdos_100} and Fig.~\ref{fig_sk_qdos_111} where the
quasiparticle bandstructure 
of the surface and central layer is plotted for a five
layer fcc(100) and fcc(111) film, respectively. Here, only the upper
subbands are shown.
While the centers of gravity of the upper $\sigma=\downarrow$ subbands 
are shifted to higher energies  for
decreasing temperatures, the lowest excitation peak in the spectral density at
$\bar\Gamma$ is even lowered due to the increasing bandwidth. 
On the other hand the width of the upper $\sigma=\uparrow$ subband decreases. 
The interplay of these two correlation effects 
leads to an inverse exchange splitting 
at the lower edge of the upper subband near to the $\bar\Gamma$ point.
The corresponding quasiparticle density of states is, however, very small.
Note, that for the same reason the position of the central peak 
of the upper subband is almost spin- and temperature-independent. 
This behavior holds for both film structures for $\vec{k}$-vectors not too 
far away from  $\bar\Gamma$.

With  help of the quasiparticle bandstructure  given in  
Fig.~\ref{fig_sk_qdos_100} and Fig.~\ref{fig_sk_qdos_111}
the exchange splitting between majority and minority spin 
direction can be analyzed in more detail. 
For both film structures the exchange splitting is 
wavevector-dependent. It is strongest near $\bar M$ for fcc(100) and between
$\bar M$ and $\bar K$ for fcc(111).   Contrary to the fcc(100) structure all
layers are fully polarized at $T=0\,$K for the fcc(111) film.
In the case of ferromagnetic saturation  
the exchange splitting can be estimated \cite{HN96} to be at most
$(1-n_{\alpha\downarrow})[4t-B_{\alpha\downarrow}]$, where 
$(1-n_{\alpha\downarrow})B_{\alpha\downarrow}$ is the effective bandshift between the 
centers of gravity of the upper quasiparticle subbands. 
For strong Coulomb interaction
$n_{\alpha\downarrow}(1-n_{\alpha\downarrow})B_{\alpha\downarrow}$ is 
proportional to the  kinetic energy \cite{HN97a} 
of the $\sigma=\downarrow$ electrons in the $\alpha$-th layer. 
Note that the kinetic energy of the  $\sigma=\uparrow$ electrons vanishes for
the ferromagnetic saturated state since the $\sigma=\uparrow$ band is completely
filled.
We want to point out that these results
strongly contrast the findings  of Hartree-Fock theory where the
exchange splitting is wavevector-independent and proportional to $m_\alpha U$
leading to substantially higher Curie temperatures compared to the MAA.
The temperature-dependence of the electronic structure within the MAA is
completely different to the Stoner picture of ferromagnetism.

Let us discuss the quasiparticle lifetime which corresponds to the width of the
quasiparticle peaks. From the spectral density 
(Figs.~\ref{fig_sk_g_qdos}, \ref{fig_sk_qdos_100}, \ref{fig_sk_qdos_111})
one can clearly read off that the
lifetime of the quasiparticles is strongly spin-  and
temperature-dependent.  
For low temperatures the upper minority spectrum is
sharply peaked
which indicates long living quasiparticles. This is due to the fact that in the
ferromagnetic saturated state a $\sigma=\downarrow$ electron does meet a
$\sigma=\uparrow$ electron at any lattice site and 
thus effectively does not perform any scattering process.  
The width of the $\sigma=\uparrow$ 
quasiparticle peaks, however,  is broadened for decreasing temperature. 
Thus in the majority spectrum the quasiparticle lifetime 
decreases for increasing magnetization. 
What concerns the lower subbands (see Fig.~\ref{fig_sk_g_qdos}) 
the respective spectrum is strongly damped
and the different excitations
due to the five layer structure are almost indistinguishable.  

For given spin and wavevector the positions of the quasiparticle peaks 
are layer-independent. In principle, their number corresponds to the number of
layers of the film. However, due to symmetry some peaks are left out for
certain layers. For thicker films the different peaks move closer 
together as their number increases until they build a continuum for 
$d\rightarrow\infty$ which corresponds to the projection of the three 
dimensional bandstructure onto the surface Brillouin zone.
Between $\bar M$ and $\bar X$ for fcc(100) and at $\bar K$ for fcc(111) the
different peaks merge together due to vanishing interlayer hopping  
($\gamma_{\perp}(\vec{k})=0$).

In Figs.~\ref{fig_sk_qdos_100},~\ref{fig_sk_qdos_111}  the QDOS of the surface
and central layer are shown 
additionally. The Van Hove singularities resulting from the different branches
of the quasiparticle dispersion are clearly visible. 
There are sharp Van Hove singularities in the minority  
spectrum while they are broadened for the majority spin direction 
because of the finite widths of the $\sigma=\uparrow$ quasiparticle peaks 
due to the quasiparticle damping.

Finally we want to stress that 
the results presented above do not depend on the 
size of the Coulomb interaction $U$ 
as long as $U$ is chosen from the strong-coupling region ($U\gg W$). 
Contrary to
Hartree-Fock theory, all magnetic key quantities  like the Curie temperature 
and the exchange splitting saturate as a function of $U$. 
On the other hand, although the MAA was optimized with respect to the strong
coupling limit we believe that, at least qualitatively, the correlation effects
in the spin-, layer-, and temperature-dependent electronic structure are valid
down to intermediate Coulomb interaction as well.

\section{Conclusion}\label{sec_conclusion}
For the investigation of spontaneous ferromagnetism and electron correlation
effects in thin itinerant-electron films 
we have applied a generalization of the modified alloy analogy (MAA) 
to the single-band Hubbard model  with reduced
translational symmetry. The MAA is based on the alloy analogy concept and is
optimized with respect to correct strong coupling behavior \cite{HL67,EO94}.
Within the MAA  the actual type of the underlying alloy 
is not predetermined  but
has to be determined selfconsistently. In this sense the MAA is able to 
account for the itineracy of the $-\sigma$ electrons which are considered as
strictly ``frozen'' in the conventional alloy analogy (AA).
In the paramagnetic phase MAA and AA are almost identical. 
However, contrary to
the AA   spontaneous ferromagnetic order is possible 
for special parameter constellations within the MAA.
With help of the MAA the interplay of magnetism 
and quasiparticle damping effects can be
studied in a natural way.

For an fcc(100) and an fcc(111) film geometry  the layer-dependent
magnetizations have been discussed as a function of temperature as well as film
thickness. The magnetization  in the surface layer is found to be reduced with
respect to the inner layers for all thicknesses and temperatures considered. 
While this reduction is weak for fcc(111) films 
it is pronounced  in the case of an
fcc(100) geometry. The effect of the surface is considerably stronger for 
fcc(100) films due to the higher percent of missing nearest neighbor atoms. 
The reduction of the surface layer magnetization is not 
to be expected within an
Hartree-Fock type approach (Stoner criterion) 
to the Hubbard film  being, therefore, a
genuine effect induced by strong electron correlations.

The magnetic behavior of the thin film systems 
can be microscopically understood by means
of the spin- layer- and temperature-
dependent quasiparticle bandstructure and the corresponding
quasiparticle density of states. There appear two correlation induced band
splittings in the quasiparticle spectrum.  
Besides the Hubbard splitting there is 
an additional exchange splitting for temperatures
below $T_C$. The demagnetization process as a function of temperature is
dominated by two distinct correlation effects: A Stoner-like
shift in the centers of gravity of the majority and minority subbands   
together with a strong spin-dependent transfer of spectral weight 
between the upper and lower subbands. 
An interplay of these two effects results
in Curie temperatures far below the corresponding Hartree-Fock values. 
The exchange splitting is found to be strongly wavevector-dependent and is
substantially different for the various quasiparticle branches in the
bandstructure. The widths of the quasiparticle peaks that correspond to the
quasiparticle lifetime exhibit a strong spin- and temperature-dependence.
For $T=0\,$K  the minority-spin  quasiparticle peaks are 
sharply peaked while the majority-spin spectrum is substantially broadened.

Clearly the degeneracy of the 3d-bands  has to  be included if a direct
comparison to the experiment is intended. Within the present scheme this could
be achieved by a similar approach as presented in \cite{NBDF89} which is
planed for the future.  However we believe the correlation effects found
here to be important within a generalized Hubbard model as well.
In this work we have exclusively focused on purely ferromagnetic films. In
addition one can examine within the same theory a phase with 
antiferromagnetic order between the layers. We expect such a situation to 
exist close to half-filling ($n=1$) and 
for intermediate values of the Coulomb interaction. 
Further the influence of a non-magnetic top layer on the magnetic behavior of
thin films  can be investigated.

\ack
This work has been done within the Sonderforschungsbereich 290 ("Metallische
d\"{u}nne Filme: Struktur, Magnetismus und elektronische Eigenschaften") of the
Deutsche Forschungsgemeinschaft.


\end{document}